\documentclass[10pt,conference]{IEEEtran}

\usepackage[utf8]{inputenc} 
\usepackage[T1]{fontenc}    
\usepackage{hyperref}       
\usepackage{url}            
\usepackage{booktabs}       
\usepackage{nicefrac}       
\usepackage{microtype}      
\usepackage[table]{xcolor}  
\usepackage{comment}
\usepackage{xspace}
\usepackage{listings,multicol}
\usepackage[caption=false]{subfig}  
\usepackage[export]{adjustbox}
\usepackage{amsmath}
\usepackage{amssymb}
\usepackage{textcomp}
\usepackage{breakurl}
\usepackage{algorithm}
\usepackage[noend]{algpseudocode}
\usepackage{cleveref}
\usepackage{multirow}
\usepackage{enumerate}
\usepackage[shortlabels]{enumitem}
\usepackage{makecell}
\usepackage{spverbatim}
\usepackage{fancyvrb}
\usepackage{tabularx}
\usepackage{tablefootnote}
\usepackage{graphicx}
\usepackage{soul}
\usepackage{balance}
\usepackage{comment}

\title{Finding Important Stack Frames in Large Systems}

\author{\IEEEauthorblockN{Aleksandr Khvorov\IEEEauthorrefmark{4}\IEEEauthorrefmark{1}, Yaroslav Golubev\IEEEauthorrefmark{2}, Denis Sushentsev\IEEEauthorrefmark{1}}
\IEEEauthorblockA{\IEEEauthorrefmark{1}\textit{JetBrains}, \IEEEauthorrefmark{2}\textit{JetBrains Research}, \IEEEauthorrefmark{4}\textit{Constructor University} \\
\{aleksandr.khvorov, yaroslav.golubev, denis.sushentsev\}@jetbrains.com
}
}

\begin{document}

\maketitle

\begin{abstract}

In this work, we developed, integrated, and tested a feature that automatically highlights potentially important frames in stack traces. The feature was implemented in the internal bug-processing tool at JetBrains that processes tens of millions of stack traces. We surveyed 18 developers at JetBrains who provided valuable feedback on the idea and the implementation.

\end{abstract}

\section{Introduction, Background, \& Existing System}

Large public-facing software systems attract a lot of bug reports from users, and efficiently processing them is crucial for the success of the products~\cite{dhaliwal, duplikates2008}. Often, the users do not write detailed reports with explicit problems, and the developers have to work with \textit{stack traces}, \textit{i.e.}, ordered lists of method calls (\textit{stack frames}) that led to the error~\cite{modani2007, bartz, karasov2022aggregation, shibaev2024stack}.

JetBrains is a large vendor of tools for software developers and teams, including IDEs such as IntelliJ IDEA~\cite{idea} and PyCharm~\cite{pycharm}. With millions of users, our company receives tens of millions of stack traces that need to be processed. To support this, JetBrains has an internal tool that collects and deduplicates incoming stack traces, as well as provides an interface for developers to study them in detail.

The default process, as it was before this work, can be seen on the left side of Figure~\ref{fig:fig}. After the developer is assigned to the issue~\cite{sushentsev2022dapstep}, they open the stack trace to study it \textbf{(a)}. The tool specifies subsystems (in green) and provides some metadata, but, importantly, does not bring one's attention to any frame in particular. During their analysis, developers often manually select frames that they consider to be important, with the result looking as \textbf{(b)}, where the developer selected frames 1 and 2 as important. This is a crucial functionality, because it is saved and shared between the developers, facilitating faster processing of the same or a similar issue in the future. However, stack traces are often long and the error is often in the middle, so we believe that the situation could be helped by automatically pre-highlighting \textit{potentially} important frames for the developer.

\section{Approach, Survey, \& Results}

\begin{figure}[t]
    \centering
    \includegraphics[width=\columnwidth]{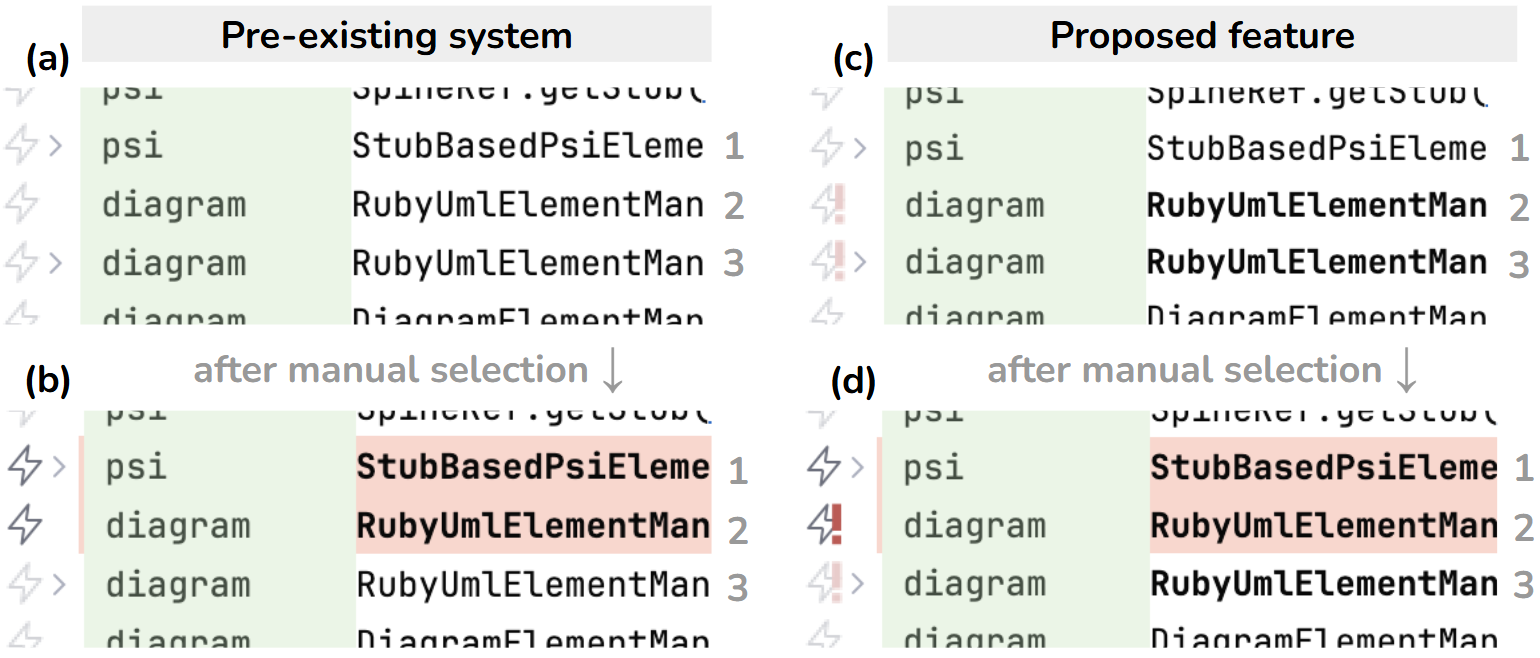}
    \vspace{-0.5cm}
    \caption{The course of work \textbf{(a-b)} in the pre-existing system without the feature, \textbf{(c-d)} with the proposed feature. The top part shows the initial UI of viewing a stack trace, the bottom --- after manually selecting important frames.}
    \vspace{-0.5cm}
    \label{fig:fig}
\end{figure}

\textbf{Approach.} The UI for the feature that we implemented can be seen on the right side of Figure~\ref{fig:fig}. Instead of an empty stack trace as in \textbf{(a)}, one can see that in \textbf{(c)}, frames 2 and 3 are pre-highlighted as potentially important with bold text and a red exclamation sign icon on the left. When the developer manually selects frames 1 and 2 \textbf{(d)}, they are highlighted exactly as before. This UI is noticeable yet at the same time unobtrusive: the suggestions are initially visible, but if the developer made their manual choice (whether selecting the suggestions or not), the suggestions do not interfere with this much more visible manual choice.

As for the way to suggest the frames, our initial version uses \textit{inverse document frequency (IDF)}~\cite{lerch} over the full corpus of received stack traces. In the given stack trace, three frames with the highest IDF (\textit{i.e.}, the frames that are the rarest in the corpus) are highlighted. The simple logic here is that the rarest frames might contain the most specific information about a particular stack trace. Highlighting top-3 frames instead of having a fixed threshold serves to not confuse the developer by sometimes having no suggestions and sometimes suggesting too much for a rare issue. In principle, other selection methods can be used with the same UI, incorporating product-specific information, dynamic analysis, AI models, etc. In this form, with IDF under the hood, the feature was \textit{bundled into production}, and is now displayed for all developers at the latest version.

\textbf{Survey methodology.} To evaluate our feature, we conducted a simple survey, asking developers to rate the usefulness and the visualization on 1--5 Likert scales, as well as to provide additional comments. We contacted 25 JetBrains developers from diverse products, 18 of whom confirmed that they recently used the tool and saw this feature, so we used their responses.

\textbf{Results and future work.} For the usefulness of the feature, the mean score was 3.6 out of 5. For the convenience of visualization, it was 4.0 out of 5, being mostly positive, with minor suggestions that relate to the coloring and the icons. In terms of major improvements, developers suggested having a tooltip that would teach them what this highlighting is (since it is not obvious). Also, they wanted the feature to motivate its selection in some way. One developer expressed their desire to correct the model so that it learns on the go. Thus, instead of IDF, it seems valuable to apply modern AI models~\cite{Du2023ResolvingCB}, which can provide better results and explicitly motivate their choice. We hope that this work provides insight into the problems  encountered when processing stack traces in large software systems and the key aspects of solving them.

\bibliographystyle{IEEEtran}
\balance
\bibliography{paper}

\newpage

\end{document}